**Title:** Longitudinal and Multimodal Recording System to Capture Real-World Patient-Clinician Conversations for AI and Encounter Research: Protocol

Misk Al Zahidy, MS[1]; Kerly Guevara Maldonado, MD[1]; Luis Vilatuna Andrango, MD[1]; Ana Cristina Proano, MD[1]; Ana Gabriela Claros, MD[1]; Maria Lizarazo Jimenez, MD[1]; David Toro-Tobon, MD[1,2]; Victor M. Montori, MD, MSc[1,2]; Oscar J. Ponce-Ponte, MD, MSc[1,3]; Juan P. Brito, MD, MSc[1,2]

1. Care and AI Laboratory, Knowledge and Evaluation Research Unit, Endocrinology, Diabetes, Metabolism and Nutrition, Department of Medicine, Mayo Clinic, Rochester, Minnesota
2. Division of Endocrinology, Diabetes, Metabolism, and Nutrition, Mayo Clinic, Rochester, MN
3. Derriford Hospital, University Hospitals Plymouth NHS Trust, Plymouth, UK.

**Corresponding author:**

Juan P. Brito, MD

brito.juan@mayo.edu

Endocrinology, Diabetes, Metabolism and Nutrition

Mayo Clinic, 200 First Street SW, Rochester, MN 55902.

**Word count: 3,879; Figures: 2; Tables: 2**

**ABSTRACT**

The promise of AI in medicine depends on learning from data that reflect what matters to patients and clinicians. Most existing models are trained on electronic health records (EHRs), which capture biological measures but rarely patient–clinician interactions. These relationships, central to care unfold across voice, text, and video, yet remain absent from datasets. As a result, AI systems trained solely on EHRs risk perpetuating a narrow biomedical view of medicine and overlooking the lived exchanges that define clinical encounters. Our objective is to design, implement, and evaluate the feasibility of a longitudinal, multimodal system for capturing patient–clinician encounters, linking 360° video/audio recordings with surveys and EHR data, to create a dataset for AI research. This single-site study is in an academic outpatient endocrinology clinic at Mayo Clinic. Adult patients with in-person visits to participating clinicians are invited to enroll. Encounters are recorded with a 360-degree video camera. After each visit, patients complete a survey on empathy, satisfaction, pace, and treatment burden. Demographic and clinical data are extracted from the EHR. Feasibility is assessed using five endpoints: clinician consent, patient consent, recording success, survey completion, and data linkage across modalities. Recruitment began in January 2025. By August 2025, 35 of 36 eligible clinicians (97%) and 212 of 281 approached patients (75%) had consented. Of consented encounters, 162 (76%) had complete recordings, and 204 (96%) completed the survey. This study aims to demonstrate the feasibility of a replicable framework for capturing the multimodal dynamics of patient–clinician encounters. By detailing workflows, endpoints, and ethical safeguards, it provides a template for longitudinal datasets and lays the foundation for AI models that incorporate the complexity of care.

# INTRODUCTION

The patient–clinician encounter is the central space where medicine is practiced, where problems are explored, treatment options negotiated, and care plans co-created. These encounters require attention to both biology and biography, encompassing not only laboratory results or physical findings but also the lived experience, values, and goals of patients. The quality of this interaction, how clinicians and patients listen, respond, and work together, determines whether care fits the realities of patients' lives and enables them to flourish despite illness. As the primary site where medicine unfolds, encounters should also be the primary lens through which we study, evaluate, and improve care[1,2].

The current paradigm in medical Artificial Intelligence (AI) is shifting from unimodal models, which analyze a single data type, to multimodal systems that integrate diverse inputs such as images, text, and structured data. To contribute meaningfully to patient care, these systems must be trained on data that reflect clinical encounters rather than relying solely on electronic health records (EHRs). Clinical notes, while indispensable for documentation, are written largely for billing and regulatory purposes and, as a result, emphasize biological measures while neglecting relational and contextual aspects of care[3,4]. As a result, they are a poor surrogate for the lived dynamics of encounters and limit the capacity of AI systems to develop a holistic understanding of patients that is necessary to foster patient-centered care.[5]

To capture both the biology and the biography of patients, including their experiences, values, and goals of care, future AI models must be trained on encounter data in which communication unfolds across multiple modalities.

The primary bottleneck for developing person-centered AI is the lack of high-quality, real-world, longitudinal multimodal data from clinical encounters. No scalable process currently exists to record and integrate this information at the level needed to support AI development and research. Recording real-world visits may disrupt clinical workflows, and both patients and clinicians must be comfortable with how data are captured, stored, and used. Consent processes need to address not only privacy and confidentiality but also future applications in AI research. Institutions must ensure ethical oversight, secure storage, and clarity around ownership and secondary use. With little precedent for routine, large-scale capture of encounter data, a feasibility study is essential to determine whether it can be done in a way that is acceptable, minimally disruptive, and ethically robust.[6,7]

To address this gap, we designed a feasibility study to test whether multimodal data from real-world patient-clinician encounters can be captured in a way that is practical, acceptable, and ethically sound. This protocol outlines the development of a replicable system that integrates 360° video and audio recordings with patient-reported surveys and EHR data. By focusing on feasibility endpoints, including clinician and patient consent, recording success, survey completion, and data linkage, this study aims to establish the foundation for the scalable collection of encounter data. Demonstrating feasibility is a critical step toward creating the longitudinal, multimodal datasets needed to train and validate next-generation AI models that reflect the complexity of clinical care.

# METHODS

## 1. Study Setting and Design

This protocol describes a step-by-step methodology to design and implement a longitudinal, multimodal encounter capture system in an academic outpatient specialty clinic (Division of Endocrinology, Mayo Clinic, Rochester, Minnesota). The protocol specifies the procedures, endpoints, and analysis plan for assessing the implementation of the encounter capture system.

Endocrinology was chosen because visits often involve chronic conditions that require longitudinal management, complex decision-making, and sustained patient–clinician collaboration. The investigators' affiliation with the division further facilitated early approvals and integration with practice. However, this protocol can be implemented across multiple healthcare services.

## 2. System Development

### 2.1 Multidisciplinary Design Process

A multidisciplinary team, including encounter-researchers, endocrinologists, implementation scientists, AI researchers, and health-services researchers, uses a rapid, iterative design approach to develop a system capable of capturing rich, multimodal data from routine clinical encounters.

### 2.2 Core Data Components

This system captures information from three integrated sources: 1) multimodal video recordings of clinical encounters, 2) post-encounter patient surveys assessing satisfaction, relational empathy, and burden, and 3) structured and unstructured EHR data, including clinical notes, patient portal messages, and clinical variables.

## 3. System Components and Workflow

### 3.1 Video Recordings

Encounters are recorded using a 360° video, 2D monocular camera (e.g., Insta360 X4) of at least 5.7K resolution and 30 frames per second (fps) (100 Mbps) to prevent camera overheating during long encounters and ensure stable recording. The camera is positioned on a discreet desktop stand in the exam room. The choice of consumer-grade hardware ensures the system records at a wider angle (360°) while keeping the system low-cost and replicable. The protocol incorporates specific mitigation strategies to address known technical limitations of this hardware.

The camera is positioned to keep participants at an optimal distance (at least 2.5 feet) and away from the camera's stitch line, a zone where visual artifacts are most common in 360° video. Audio is captured in dual channel at a 48 kHz sample rate using the camera's "Stereo" mode ensuring both patient and clinician voices with equal fidelity, which can be down sampled to 16kHz for speech analysis. "Wind Reduction" settings are avoided as they can distort vocal frequencies critical for analysis. To manage thermal load and ensure continuous operation, batteries are swapped between each recorded encounter. The 5.7K/30fps setting allows for at least 135-minute recording time in the latest 360°, 2D monocular camera, sufficient for typical outpatient visits. Participants are informed that they can pause, redirect, or disable recording at any time. The daily workflow for encounter capture is illustrated in **Figure 1.**

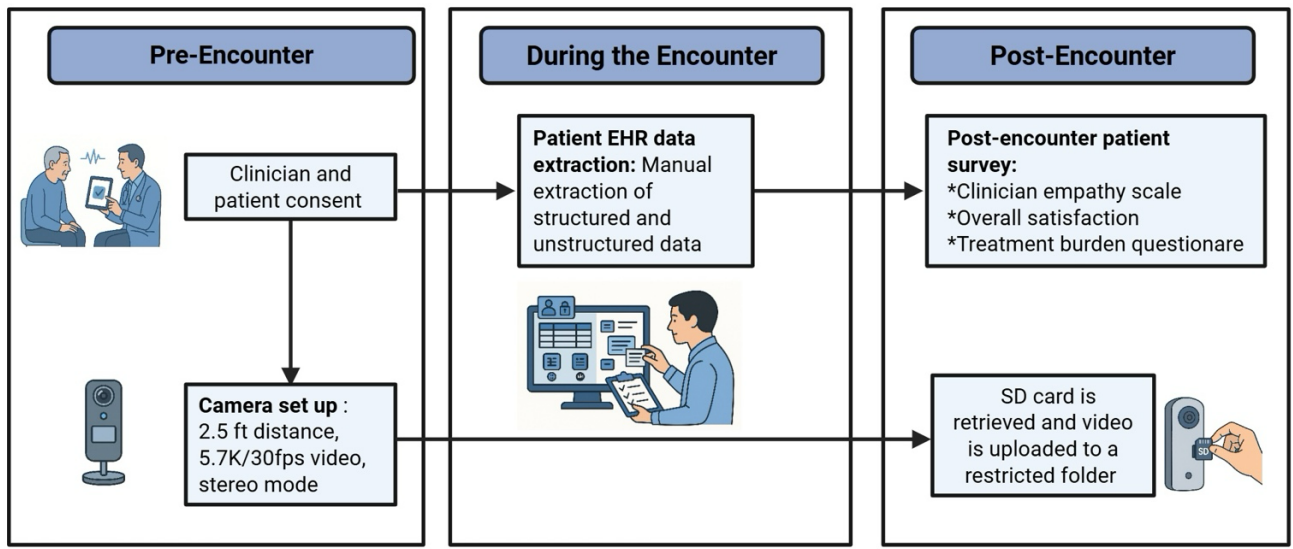

**Figure 1. Workflow of multimodal encounter capture. The system integrates three stages: pre-encounter (clinician and patient consent, camera setup), during-encounter (video/audio recording and manual EHR data extraction), and post-encounter (patient survey completion and secure video upload).**

A standardized workflow is followed daily to ensure data integrity and security. At the end of each clinic session, the camera's Secure Digital (SD) card is immediately retrieved and files saved in their corresponding format, converted to mp4 when needed, and uploaded to a restricted-access, HIPAA-compliant, institutional server. All files are renamed using the harmonized participant ID and naming convention. Subsequent encounters reuse the same participant ID with a new encounter suffix. Once the upload is verified, the raw and processed files are permanently deleted from the SD card and local workstation.

3.2 Post-Encounter Patient Surveys

Immediately following the clinical encounter, patients are invited to complete a brief post-visit survey administered on a tablet. The survey includes validated scales to assess clinician empathy, overall satisfaction, visit pace, Treatment Burden Questionnaire (TBQ) and perceived burden of the visit[8-12] **(Appendix A).** It also includes two open-ended questions designed to elicit qualitative feedback:

- "Tell us how, if at all, did your visit with an endocrinology clinician help you?"
- "If you had a magic wand and you could change one thing about your visit today, what would that be?"

If a patient is unable to complete the survey in clinic, study staff follow up with up to three reminders via email and patient portal messages.

### 3.3 EHR Data Extraction

Following a recorded encounter, trained study staff perform a manual extraction of structured (demographic and clinical variables) and unstructured (clinical notes, patient portal messages) data from the patient's EHR. Structured data include age, sex, race/ethnicity, clinical diagnoses, visit type (e.g., new or follow-up), and current medications.

### 4. Pilot Phase and Iterative Refinement

The system is piloted over three weeks, with 30 recorded encounters (~10 per week) to identify and resolve logistical issues related to camera placement (e.g., ensuring the camera is unobtrusive), workflow disruption (e.g., additional steps or delays during rooming), and recording quality (e.g., interruptions or file-transfer failures). Pilot feedback is reviewed weekly, and the recruitment script, equipment handling, and data processing steps are refined before initiating full-scale recruitment. Recorded encounters from the pilot phase can be included in the final dataset. The system is refined based on three design constraints to ensure the final system is scalable and replicable:

1. Comprehensive video capture: The system must provide 360-degree video coverage without requiring a dedicated camera operator in the room.

2. High-quality audio: The system must capture dual-channel audio at a sample rate sufficient for automated speech recognition and diarization ($\geq$ 16 kHz).

3. Minimal clinic disruption: Equipment setup and recording initiation must fit within the standard patient rooming process, without extending visit times.

### 5. Participant Recruitment and Eligibility Criteria

### 5.1 Clinicians

All practicing clinicians in the Division of Endocrinology, including physicians, advanced practice providers, fellows, certified diabetes care and education specialists, and registered nurses, are eligible to participate. Clinicians are approached in person or via email, provided with detailed study information, and consented electronically on a tablet. Upon consent, each clinician is assigned a unique participant ID (e.g., ENDO-C-###) and receives an email to complete a short demographic survey which includes their professional role, age, gender, race/ethnicity, and years in practice.

### 5.2 Patients

Patients are eligible if they are adults (aged 18 years or older), have a scheduled in-person appointment with a participating clinician, and can provide informed consent in English. Exclusion criteria include the presence of cognitive impairment, need for an interpreter, or not being the primary individual responsible for managing their care. Enrolled patients are assigned a unique participant ID (i.e., ENDO-P-###) for longitudinal linkage across visits.

Any adult guests present in the exam room (e.g., spouse, adult child) must provide verbal consent before recording. If a guest declines to be recorded or a minor is present, the encounter is not recorded for that visit to ensure privacy and compliance. The full process of recruitment is visualized in **Figure 2**.

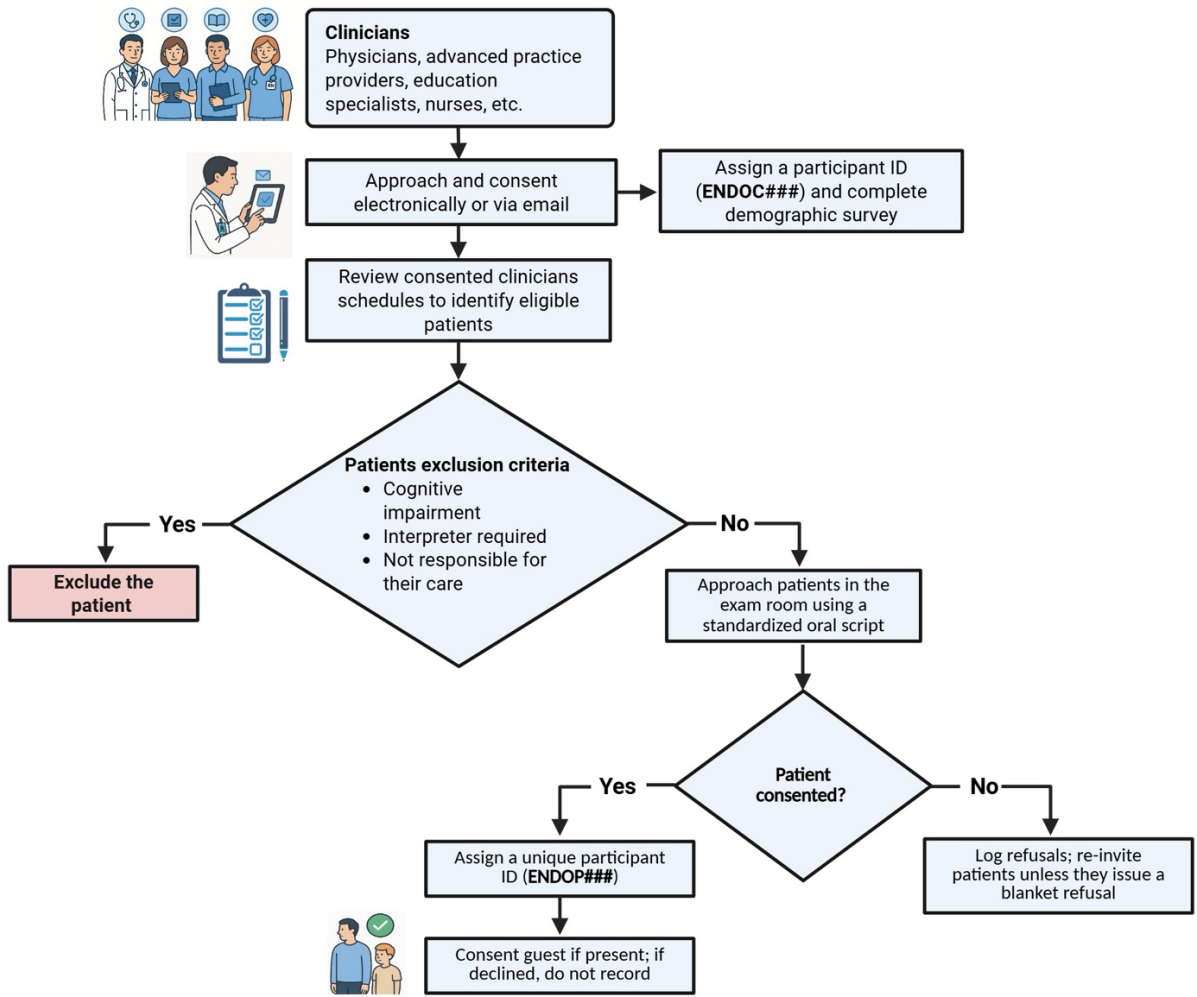

**Figure 2. Recruitment and consent workflow for clinicians and patients.**

5.3 Recruitment Process

A team of five trained non-clinical research staff manages all recruitment and data collection activities. The standardized recruitment process is as follows:

1. Daily Schedule Review: Research staff review participating clinicians' schedules to identify eligible new patients and previously consented patients returning for follow-up appointments.
2. Low-Pressure Approach: Research staff approach eligible patients in the exam room using a standardized, IRB-approved oral script **(Appendix B)**. The script uses clear, lay language to explain the study's purpose, emphasizing that participation is entirely voluntary, privacy will be protected, and the recording can be paused or stopped at any time without affecting their care.
3. Consent and Confirmation: New patients provide informed consent on a tablet. For returning patients who have previously consented, staff obtain verbal confirmation that they are still comfortable being recorded. Full written re-consent is only required if there are protocol changes.
4. Declinations: Patients may decline recording for any appointment. Reasons for refusal are logged by staff to monitor for potential selection bias. Patients who decline are not re-approached the same day but may be invited to participate again at future visits unless they issue a blanket refusal for all future participation.

6. Data Integration and Management

6.1 Triangulation and Harmonization

The study is designed to capture three components: (1) 360° video-recorded clinical interactions, (2) post-encounter patient surveys, and (3) structured and unstructured data extracts from the EHR. These sources are harmonized using an identification schema and documented in a central tracking log.

For clinicians, the log tracks their professional role, consent status, and completion of the demographic survey. For patients, it records demographic data (age, sex, and race/ethnicity), eligibility status with documented reasons for ineligibility, guest presence and consent status, recording success, video upload success, post-visit survey completion status (complete, partial, or none), and any challenges noted by the research staff (e.g., equipment malfunction, patient discomfort). Each entry is indexed by the unique participant ID, connecting all three data streams. All files follow a harmonized naming convention based on the participant ID and encounter date to allow reliable data merging and adherence to FAIR (Findable, Accessible, Interoperable, Reusable) data principles to maximize the utility of the collected data for future research.

**Table 1** outlines the data sources, formats, linkage keys, and storage platforms. This structured approach ensures reliable data merging and supports complex, temporally aligned multimodal analyses.

**Table 1: Multimodal Data Streams, Key Variables, and Linkage Schema**

| Data Source | Modality | Key Variables / Format | Linkage Key | Storage Platform |
|---|---|---|---|---|
| **Camera** | Video | 360° video, 5.7K/30fps, H.265 codec, .mp4 format | PATIENT_ID_ENCOUNTER_DATE.mp4 | Secure Institutional Server |
| **Camera** | Audio | Dual-channel, 48 kHz WAV, down sampled to 16 kHz | PATIENT_ID_ENCOUNTER_DATE.wav | Secure Institutional Server |
| **Survey** | Patient-Reported | Empathy, satisfaction, burden (Likert scales); Open-ended text | PATIENT_ID, ENCOUNTER_DATE | REDCap |
| **EHR** | Structured and Unstructured information | Demographics, ICD-10 codes, medications, lab values | PATIENT_ID, ENCOUNTER_DATE | REDCap (via manual extraction) |

6.2 Storage and Security

All study data are stored in accordance with HIPAA regulations and institutional data security protocols. Video and audio recordings are labeled with study-specific IDs and uploaded daily to a secure, password-protected institutional server with access restricted to authorized members of the core research team.

Survey and EHR data are managed and stored using REDCap, a secure, web-based application designed for research data collection and management. The final analysis dataset is fully de-identified, with the unique participant ID serving as the only link between the different data modalities. Patients who withdrew consent or became ineligible during the study are excluded from the final dataset unless explicit permission is granted to retain their previously collected data.

7. Data Analysis

This protocol includes feasibility and recruitment results; analysis of the study components will be reported in a later publication.

7.1 Feasibility and Recruitment Metrics

The study's feasibility is measured using five predefined endpoints. These endpoints, their definitions, and the a priori success thresholds are detailed in Table 2. Descriptive statistics (counts and proportions) are calculated for each endpoint using data from the recruitment tracking logs and REDCap exports.

**Table 2: Primary Feasibility Endpoints, Definitions, and Success Thresholds**

| Endpoint | Definition (Numerator / Denominator) | Success Threshold |
|---|---|---|
| *1. Clinician Consent Rate* | Consented clinicians / All approached clinicians | > 90% |
| *2. Patient Consent Rate* | Consented patients / All approached eligible patients | > 50% |
| *3. Recording Success Rate* | Full-length recordings / All consented encounters | > 90% |
| *4. Survey Completion Rate* | Completed surveys / All consented encounters | > 90% |
| *5. Data Linkage Rate* | Encounters with linked video, survey, & EHR / All recorded encounters | > 90% |

7.2 Planned Analysis of Data Components

The analysis of the three components (video recordings, post-encounter surveys, and EHR data) will be described, in a future publication, using descriptive statistics.

*Acoustic feature extraction:*

Video recordings will be automatically transcribed and diarized using the best available models.[13,14] From the audio, key acoustic features will be extracted using the Parselmouth Python library, which provides access to Praat's gold-standard algorithms. These include: 1) prosodic features (fundamental frequency mean and standard deviation, intensity contours), 2) spectral features (formants F1-F3, related to vowel articulation, and 3) Temporal features (speech rate, pause duration and frequency). This will allow for quantitative analysis of conversational dynamics such as turn-taking, speech overlap, and backchannels. Visual features such as smile, nod, gaze, and emotion detection will be extracted with computer vision tools such as OpenCV and Emotion-Qwen.[15,16]

*Visual feature extraction:*

A planned analysis pipeline using the Open Computer Vision Library (OpenCV)[15] will be employed to extract nonverbal cues. This process will involve: 1) Face detection using pre-trained deep neural network models to locate patient and clinician faces in each frame[15,17], 2) Facial landmark detection to identify key points (e.g., corners of the eyes and mouth)[15,18], and 3) approximations of Facial Action Units (AUs) corresponding to specific muscle movements (e.g., smiling, brow furrowing)[15,18].

*Advanced emotion and behavior analysis:*

The dataset will be leveraged to train and validate advanced models like Emotion-Qwen[16], a state-of-the-art Large Multimodal Model. This will enable analysis that moves beyond simple emotion labels (e.g., 'happy', 'sad') to more complex, context-aware reasoning about the affective dynamics of the conversation, such as identifying moments of empathy, confusion, or shared understanding.

*Post-encounter surveys and EHR data:*

Data from the post-encounter surveys will be used to calculate the mean scores, standard deviations, and distribution of each scale item. Where relevant, responses will be stratified by key variables of interest, such as clinician type (e.g., physician, fellow, educator) and visit type (e.g., new vs. follow-up). For the subset of patients who complete the TBQ, total scores and item-level responses will be summarized to describe perceived treatment workload. Analysis will be performed using SAS (version 9.4; SAS Inst., Cary, NC, USA).

Structured data from EHRs will be described in the same manner as post-encounter survey results whereas unstructured data (i.e., clinical notes and patient portal messages) will be described in terms of length (number of words).

8. Ethical Considerations

This study was approved by the Mayo Clinic Institutional Review Board (IRB #24-012956). All clinicians, patients, and adult guests give informed consent prior to recording. Recordings and linked data are stored in secure, HIPAA-compliant institutional servers. Participants may pause or withdraw at any time without affecting their clinical care.

**RESULTS**

The protocol was successfully implemented in the outpatient clinic of the Division of Endocrinology at Mayo Clinic in Rochester, Minnesota, an academic specialty clinic. We selected the Insta360 X4 camera because it met all required specifications, including 5.7K resolution at 30 frames per second (~100 Mbps) with dual-channel 48 kHz audio, for a longitudinal 360° 2D encounter capture system. Video files were recorded in INSV format and converted to MP4 using the Insta360 software. Pilot-phase recordings were included in the final dataset, as they were deemed to be of sufficient quality by the research team.

Recruitment and encounter recording began in January 2025. By August 2025, 35 of 36 eligible clinicians (97%) and 212 of 281 approached patients (75%) had consented. Of the consented encounters, 162 (76%) resulted in a complete and usable 360° video recording, and 204 patients (96%) completed the post-visit survey. Final analyses are planned with feasibility findings anticipated for publication in early 2026. Preliminary results indicate that the recruitment strategy and data collection workflow are viable.

**DISCUSSION**

1. Principal Results

This protocol describes a scalable framework for capturing rich, multimodal, longitudinal, real-world clinical encounters in chronic disease care. The primary anticipated outcome of this study was to demonstrate feasibility, determined by achieving predefined thresholds for recruitment, data capture, and data linkage. Preliminary findings indicate that these thresholds are being met: 97% of eligible clinicians and 75% of approached patients consented, 76% of encounters resulted in a complete and usable 360° video recording, and 96% of patients completed the post-visit survey. These early indicators support the practicality of this approach in an outpatient specialty clinic and provide a foundation for subsequent analysis. The collected corpus addresses critical gaps in person-centered AI research by providing the raw material needed to train and validate next-generation clinical AI models that can understand and respond to the full context of patient care.

2. Comparison with Prior Work

There have been similar efforts to capture multimodal information from natural conversations, but few have focused on real-world healthcare encounters. To the best of our knowledge, prior studies in healthcare have been limited to audio-only transcript analysis like applying NLP to primary care conversations[6], or video capture in simulated training settings rather than routine clinical care.[19] Outside of healthcare, multimodal datasets such as the CANDOR corpus have advanced the study of natural conversation, but these do not address the ethical, logistical, and workflow challenges unique to clinical environments, such as HIPAA compliance and the need to avoid disrupting care delivery.[7]

This protocol advances the field by embedding multimodal capture directly into routine specialty care, an approach not addressed in previous studies. The iterative piloting with clinical staff illustrates stakeholder engagement, ensuring the system fit existing workflows and minimized disruption, a principle emphasized in implementation science. The harmonized data linkage strategy enhances feasibility and replicability by reliably integrating video, survey, and EHR data. Finally, the use of low-cost, general-purpose hardware increases scalability and democratizes access to multimodal encounter research, including resource-limited settings.

3. Next Steps and Future Directions

The immediate next steps involve analyzing the rich dataset this protocol enables, with initial work focused on correlating patient-reported outcomes with observable communication dynamics. For example, verbal and nonverbal behaviors such as speaking time, turn-taking, and gaze can be examined in relation to patient-reported empathy, satisfaction, and treatment burden. These analyses aim not only to document burden but also to identify conversational moments that foster patient-centeredness such as instances where empathy, attentiveness, or shared decision-making are enacted despite the constraints of industrialized healthcare. By linking such moments to patient-reported outcomes, we can begin to specify communication practices that either alleviate or exacerbate a patient's perceived burden of care, particularly in relation to navigating services and managing costs.

Beyond these initial projects, the dataset presents significant opportunities for advanced AI research, though not without challenges. Processing each modality introduces unique complexities: audio data requires robust speaker diarization and transcription models tuned for clinical conversations; video data presents challenges in data volume, storage, and the computational cost of extracting nonverbal cues like facial expressions and gestures; and integrating these streams with EHR and survey data requires sophisticated data fusion techniques to handle issues like missing data and ensure temporal alignment.

Furthermore, while this protocol was implemented in endocrinology, the framework is designed for broader applicability. Future work will involve deploying this system in other clinical settings, such as oncology, where complex treatment decisions and longitudinal patient relationships are also central to care. In oncology, integrating encounter data with genomic, imaging, and clinical data could train AI models to support shared decision-making and personalize patient communication, augmenting rather than replacing clinical judgment. By systematically addressing these next steps, this work can help build the foundational datasets needed to move beyond EHR-based models and develop AI that truly understands the human context of clinical care.

Finally, it is important to address the question of scalability. While this protocol describes a replicable framework, the long-term vision is not to place a camera in every exam room for every encounter, which may not be practical or desirable. Instead, this foundational research is a crucial step toward the development of more integrated and less obtrusive ambient intelligence technologies. These future systems, embedded seamlessly into the clinical environment, could capture the essential multimodal data of an encounter without the need for standalone recording devices. The insights gained from this study, regarding consent workflows, patient and clinician acceptance, and ethical safeguards, will be invaluable for informing the design of next-generation ambient systems that can capture encounter data at scale while maintaining the trust and privacy essential to the patient-clinician relationship. By systematically addressing these next steps, this work can help build the foundational datasets needed to move beyond EHR-based models and develop AI that truly understands the human context of clinical care.

4. Limitations

This framework faces potential challenges, and the protocol includes specific strategies to mitigate them. One concern is selection bias, where patients who consent to participate may differ systematically from those who do not, for example, by being more comfortable with technology or having higher trust in the healthcare system. To address this, the protocol includes logging reasons for refusal, using a low-pressure consent script to build trust, and conducting a comparative analysis of key variables between participants and non-participants to quantify any observed bias. Another potential issue is the Hawthorne effect, where the presence of a recording device may alter the natural behavior of participants. The study's longitudinal design serves as the primary mitigation, as reactivity is expected to diminish over time as participants become habituated to the recording process. The use of a small, unobtrusive camera is also intended to minimize the salience of the observation.

The protocol also proactively addresses potential challenges related to the technical reliability of consumer-grade hardware. Because high-resolution 360° video can overheat cameras and exhaust storage, our protocol includes capping resolution at 5.7 K, rotating batteries on a fixed schedule, and formatting SD cards daily. Finally, as a single-site study in a specialized clinic, the findings may not be immediately generalizable. This study is intended as a foundational step, and if feasibility is demonstrated, future multi-site trials will be essential to establish broader applicability.

5. Conclusions

This protocol describes a rigorous, transparent, and replicable method for capturing the multimodal dynamics of clinical encounters in chronic disease care. By detailing operational workflows, technical specifications, feasibility endpoints, and mitigation strategies, it offers a practical template for other research teams. Demonstrating the feasibility of this longitudinal approach using low-cost equipment will create a valuable resource for developing next-generation, person-centered AI tools that can understand and respond to patients lived experiences. This work represents a foundational step toward building AI that can appreciate both the "biology" and the "biography" of a patient, helping make healthcare more effective, equitable, and humane.

**ACKNOWLEDGEMENTS**

We would like to acknowledge the desk operations specialists and clinicians in the Division of Endocrinology for their support and collaboration during the design and implementation of this study. Their assistance in coordinating clinic logistics and facilitating recruitment was essential to the success of this protocol.

**DATA AVAILABILITY**

The data generated in this study include identifiable patient–clinician encounter recordings and linked electronic health record data. Due to privacy and ethical restrictions, these data cannot be made publicly available. De-identified data may be made available upon reasonable request to the corresponding author and with appropriate institutional approvals.

**AUTHORS' CONTRIBUTIONS**

MAZ, OJPP, DTT, and JPB conceived the study. MAZ led protocol drafting, system design, and preparation of the manuscript. KGM, LVA, ACP, and AGC contributed to patient recruitment, data collection, and drafting sections of the manuscript. MLJ assisted with manuscript preparation and figure development. JPB provided overall supervision, clinical oversight, and critical revision of the manuscript. All authors reviewed and approved the final version.

**CONFLICTS OF INTEREST**

The authors declare that they have no conflicts of interest related to this work.

**ABBREVIATIONS**

- AI: Artificial Intelligence
- CARE: Consultation and Relational Empathy
- EHR: Electronic Health Record
- HIPAA: Health Insurance Portability and Accountability Act
- IRB: Institutional Review Board
- NPS: Net Promoter Score
- SD: Secure Digital
- TBQ: Treatment Burden Questionnaire
- UCAT: Unhurried Conversations Assessment Tool

**Appendix A. Post-Encounter Patient Survey**

# DID WE CARE WELL TODAY?
A research study

Thank you for helping us with this study. We are asking you about your visit today. Your responses are completely confidential and voluntary, will be used for research only, and will not be shared with anyone outside of the research team.

**What is the main reason for visiting Endocrinology at Mayo Clinic today?** *(Select more than one if needed)*
☐ Thyroid Nodule
☐ Thyroid Cancer
☐ Other Thyroid Condition
☐ Diabetes
☐ Weight Management
☐ Osteoporosis or Osteopenia
☐ Adrenal Disorders
☐ Pituitary Disorders
☐ Calcium or Parathyroid Concerns
☐ Testosterone Concern
☐ Polycystic Ovary Syndrome (PCOS)
☐ Other (please specify): ______________________
☐ Prefer Not to Disclose

**Is this your first visit to Endocrinology at Mayo Clinic for this problem?**

☐ Yes
☐ No
☐ Not sure
☐ Prefer not to answer

**Kindly mark the box next to each statement to indicate how was your endocrinology visit was today.**

| How good was the clinician at... | Poor | Fair | Good | Very Good | Excellent | Does not apply |
|---|---|---|---|---|---|---|
| **1) Making you feel at ease** (introducing themselves, being friendly, respectful, and warm towards you; not cold or abrupt) | ☐ | ☐ | ☐ | ☐ | ☐ | |
| **2) Letting you tell your "story"** (giving you time to describe your condition in your own words; not interrupting, rushing or diverting you) | | | | | | |
| **3) Really listening** (paying close attention to what you were saying, not looking at the notes or computer as you were talking) | | | | | | |
| **4) Being interested in you as a whole person** (asking/knowing relevant details about your life, your situation; not treating you as "just a number") | | | | | | |
| **5) Fully understanding your concerns** (communicating that they had accurately understood your concerns and anxieties; not overlooking or dismissing anything) | | | | | | |
| **6) Showing care and compassion** (seeming genuinely concerned, connecting with you on a human level; not being indifferent or "detached") | | | | | | |
| **7) Being positive** (having a positive approach and a positive attitude; being honest but not negative about your problems) | | | | | | |
| **8) Explaining things clearly** (fully answering your questions; providing clear and adequate information; not being vague) | | | | | | |
| **9) Helping you to take control** (exploring with you what you can do to improve your health yourself; encouraging rather than "lecturing" you) | | | | | | |
| **10) Making a plan of action with you** (discussing the options, involving you in decisions as much as you want to be involved; not ignoring your views) | | | | | | |

**Kindly mark the box next to each statement to indicate how your endocrinology visit was today.**

|  | Not very | Minimally | Unsure | Somewhat | Very |
|---|---|---|---|---|---|
| **11)** How **helpful** was your visit with the clinician? | | | | | |
| **12)** How **rushed** was your visit with the clinician today? | | | | | |
|  | Not at all | Probably not | Unsure | Probably yes | Defnitely |
| **13)** Likelihood of you recommending this clinician to others | | | | | |

**Tell us how, if at all, did your visit with an endocrinology clinician help you?**

**Now take a moment and consider everything you have to do to take care of your health. Please rate the burden or problem associated with each of the following items using the following scale:**

**How would you rate the problems related to:**

Arranging medical appointments (doctor visits, lab tests and other exams) and reorganizing your schedule around these appointments

| Does not apply | Not a Problem 0 | 1 | 2 | 3 | 4 | 5 | 6 | 7 | 8 | 9 | Big Problem 10 |
|---|---|---|---|---|---|---|---|---|---|---|---|
| ☐ | ☐ | ☐ | ☐ | ☐ | ☐ | ☐ | ☐ | ☐ | ☐ | ☐ | ☐ |

The financial burden associated with your healthcare (for example: out of pocket expenses or expenses not covered by insurance)?

| Does not apply | Not a Problem 0 | 1 | 2 | 3 | 4 | 5 | 6 | 7 | 8 | 9 | Big Problem 10 |
|---|---|---|---|---|---|---|---|---|---|---|---|
| ☐ | ☐ | ☐ | ☐ | ☐ | ☐ | ☐ | ☐ | ☐ | ☐ | ☐ | ☐ |

**If you had a magic wand and you could change one thing about your visit today, what would that be?**

```
[                                                                            ]
```

**Appendix B. Recruitment Script**

*Hello, my name is [NAME]. I am part of the research team here in the [Department/Clinic]. We are conducting a research study to better understand how patients and clinicians communicate and interact during visits. Participation is voluntary. If you are interested, you would sign a consent form, we would place a small recording device in the room during your appointment, and after your visit you would complete a short survey. The device records video and audio of the visit, but most people forget about it after a few minutes. You can pause or stop the recording at any time, and your care will not be affected if you choose not to participate. Would you like to hear more about the study?*